\documentclass[11pt,twoside]{article}
\usepackage{asp2010}

\resetcounters

\begin{document}

\title{Habitability of Planets Orbiting Cool Stars}
\author{Rory Barnes$^{1,2}$, Victoria S. Meadows$^{1,2}$, Shawn D. Domagal-Goldman$^{2,3,4}$, Ren{\' e} Heller$^5$, Brian Jackson$^{4,6}$, Mercedes L{\' o}pez-Morales$^{7,8}$, Angelle Tanner$^9$, Natalia G{\' o}mez-P{\' e}rez$^{8,10}$, Thomas Ruedas$^8$}
\affil{$^1$Astronomy Department, University of Washington, Box 351580, Seattle, WA 98195, USA}
\affil{$^2$Virtual Planetary Lab, USA}
\affil{$^3$Planetary Sciences Division, NASA Headquarters, 300 E
St. SW, Washington, DC, USA}
\affil{$^4$NASA Postdoctoral Program Fellow}
\affil{$^5$Astrophysikalisches Institut Potsdam (AIP), An der Sternwarte 16, 14482 Potsdam, Germany}
\affil{$^6$NASA Goddard Space Flight Center, Greenbelt, MD 20771, USA}
\affil{$^7$Institut de Ci\`encies de L'Espai (CSIC-IEEC), Campus UAB,
 Fac. Ci\`encies. Torre C5 parell 2, 08193 Bellaterra, Barcelona, Spain}
\affil{$^8$Carnegie Institution of Washington, Dept. of Terrestrial Magnetism, 5241 Broad Branch Road NW, Washington, DC, 20015, USA}
\affil{$^9$Department of Physics and Astronomy, Georgia State
University, One Park Place, Atlanta, GA 30303, USA}
\affil{$^{10}$Departamento de F{\' i}sica, Universidad de los Andes, Cr 1E No 18A-10, Bogot{\' a}, Colombia.}

\begin{abstract}
Terrestrial planets are more likely to be detected if they orbit M
dwarfs due to the favorable planet/star size and mass ratios.
However, M dwarf habitable zones are significantly closer to the star
than the one around our Sun, which leads to different requirements for
planetary habitability and its detection.  We review 1) the current
limits to detection, 2) the role of M dwarf spectral energy
distributions on atmospheric chemistry, 3) tidal effects, stressing
that tidal locking is not synonymous with synchronous rotation, 4) the
role of atmospheric mass loss and propose that some habitable worlds
may be the volatile-rich, evaporated cores of giant planets, and 5)
the role of planetary rotation and magnetic field generation,
emphasizing that slow rotation does not preclude strong magnetic
fields and their shielding of the surface from stellar
activity. Finally we present preliminary findings of the NASA
Astrobiology Institute's workshop ``Revisiting the Habitable Zone.''
We assess the recently-announced planet Gl 581 g and find no obvious
barriers to habitability. We conclude that no known phenomenon
completely precludes the habitability of terrestrial planets orbiting
cool stars.
\end{abstract}

\section{Introduction}

Initially dismissed as potential habitats, planets orbiting M dwarfs
have lately seen renewed interest \citep{Tarter2007,Scalo2007}. With
lower luminosities, their ``habitable zones'' (HZ), the range of
orbits in which an Earth-like planet could support surface water
\citep{Kasting1993}, are significantly closer-in than for solar-type
stars. This proximity leads to new perils such as increased
susceptibility to stellar activity and stronger tidal effects. The
discovery of exoplanets spurred more detailed and careful evaluation
of these phenomena, and many researchers now argue that they are not
as dangerous for life as previously feared. On the contrary, these
planets may be ideal laboratories to test models of geophysics,
atmospheric dynamics, celestial mechanics, photochemistry, aeronomy,
and ultimately habitability.

Without analogs in our Solar System or adequate remote sensing
capabilities, the surface properties of M dwarf planets can only be
considered theoretically. Nonetheless, progress has been made in
several areas, such as modeling of planetary interiors
\citep{Sotin2007,ONeillLenardic2007}, atmospheric mass loss
\citep{Yelle2004,Segura2010}, atmospheric dynamics
\citep{Joshi2003,HengVogt2010}, and tidal effects
\citep{Jackson2008c,Heller_sub}. This chapter is a
multidisciplinary study of the potential habitability of M star
planets, but with an astrophysical bias. A full treatment would
require far more space than permitted by this format. For a more
comprehensive analysis of M dwarf planet habitability (and
habitability in general), see \citet{Tarter2007}, \citet{Scalo2007},
and $\S$ 7.

This chapter is organized as follows. First we examine the current
detection limits of terrestrial planets due to stellar
variability. Second, we explore the different chemical reactions in
planetary atmospheres due to different stellar spectral energy
distributions. Next we examine tidal effects. Fourth, we consider the
possible existence of ``habitable evaporated cores'' of giant
planets. Fifth, we explore the magnetic fields of terrestrial
planets. Finally, we report key interdisciplinary findings from a
recent NASA Astrobiology Institute workshop titled ``Revisiting the
Habitable Zone.''

\section{Is M Dwarf Variability a Barrier to Detecting Earth-Mass Planets?}

With $>500$ extra-solar giant and super-Earth planets detected around
nearby stars, we are now probing the dependence on stellar mass of
planet formation and migration. At the same time, we are pushing the
limits of radial velocity (RV) planet detection methods below
``super-Earths'' (planet mass $m_p \le 10~\textrm{M}_\oplus$) and toward a bona
fide Earth analog, e.g. \citet{Mayor2009}. M dwarfs play a critical
role in both of these scientific pursuits.  The diminutive masses of M
dwarfs result in a higher sensitivity to lower-mass planets at a given
RV precision, and many of the lightest exoplanets have been detected
around M dwarfs with semi-amplitudes of $\sim 1$ m/s
\citep{Mayor2009}. New instruments have the potential to push the
instrumental residuals down to 10 cm/s \citep{PepeLovis2008}, the
amplitude of an Earth-mass planet in the habitable zone of a
solar-type star. The amplitude of that same planet around an
early-type M dwarf is 30-60 cm/s and $>1$ m/s for late-M dwarfs. These
limits imply that we currently have the instrumental precision
necessary to detect terrestrial planets in the HZs of the nearest M
dwarfs.

On the other hand, M dwarfs are hundreds of times fainter than G stars
at optical wavelengths. As a result, there has been a surge in
near-infrared RV survey efforts with high-resolution spectrographs
such as NIRSPEC on Keck and CRIRES on the VLT. Using telluric lines
for wavelength calibration, these surveys are capable of reaching RV
precisions of 20-100 m/s on M and L dwarfs depending on the rotation
rate of the star \citep{Blake2010}. With additional calibration from
either an ammonia or methane gas cell, these instruments can reach RV
precisions of 5-10 m/s for mid to late M dwarfs
\citep{Bean2010a,Bean2010b}. While the amplitude of the RV signature
is larger around these lighter mass stars, M dwarfs suffer from more
flaring and starspot activity than their solar-type
counterparts. These photospheric activity sources introduce
perturbations, often called jitter, into both the photocenter of the
star and the disk-averaged radial velocity. The degree of the
perturbation depends on the rotation rate of the star, the flare
occurrence rate, the starspot lifetime and the degree of starspot
coverage. \citet{Makarov2009} used a simple starspot model to estimate
that RV jitter for F, G and K stars can reach up to
69, 38 and 23 cm/s, respectively and, thus, inhibit our ability to
detect Earth-mass planets in the HZ.

Stellar variability may be measured in several different ways.  The
ultra-precise photometry from the {\it Kepler} telescope shows that,
in general, early-M dwarfs display less astrophysical jitter than
their FGK counterparts \citep{Basri2010,Ciardi2010}. However, {\it
Kepler} data include few mid-M dwarfs and no late-M
dwarfs. Additionally, jitter levels may vary from star to star within
the same spectral class suggesting we need to quantify the jitter for
individual targets. In fact, in some instances the variability of the
H$\alpha$ line, a common proxy for stellar activity, is {\it
anti-correlated} with the RV jitter \citep{Zechmeister2009}.

Starspots are another source of astrophysical noise, and many groups
have developed basic (one continuous, circular spot) and complex
(various spot lifetimes and temperatures, multiple spots at a range of
latitudes) models. The expected contribution to RV jitter from
starpots on M dwarfs was addressed by \citet{Reiners2010} using a
model with a single spot and multiple wavelengths and stellar rotation
rates. For example, a 2800 K star with a 2600 K spot and a 2 km/s
rotation rate will have an RV jitter of 10 m/s in the optical and 2
m/s in the infrared. These levels go up to 14 and 10 m/s,
respectively, for a rotation rate of 10 km/s. This jitter is larger
than the 2 m/s precision needed to detect an Earth-mass planet in the
habitable zone of an M6 star.

RV technology will soon reach the 10 cm/s precision necessary to
detect Earth-mass planets in the HZs of nearby stars. However, we do
not know the levels of intrinsic stellar jitter of stars in our local
neighborhood -- the majority of which are M dwarfs. Recent advances
have demonstrated that broad assumptions based on spectral type are
inadequate. Precise photometry and complex star spot modeling, as well
as the forthcoming statistical results from the {\it Kepler, MOST} and
{\it CoRoT} missions could provide valuable insight, but, in certain
cases, jitter may prevent the detection of terrestrial planets.

\section{Implications of M Dwarf Spectral Energy Distributions on Life Detectability}

Most plans to search for and characterize life on extrasolar planets
involve probing the atmospheric chemistry of those planets by
analyzing their spectra to search for specific gaseous components
which can only be produced by biological processes. For example, there
have been calls to search for the simultaneous presence of methane
(CH$_4$) and either molecular oxygen (O$_2$), or its photochemical
by-product, ozone (O$_3$) \citep[for a review
see][]{DesMarais2002}. Others have suggested looking for life by
searching for methyl-chloride (CH$_3$Cl) \citep{Segura2005}, or
nitrous oxide (N$_2$O) \citep{Sagan1993}. In this section, we discuss
how the stellar spectral energy distribution (SED) of M dwarfs can
significantly impact our ability to identify biospheres.

Different SEDs profoundly affect atmospheric chemistry, as many
reactions are the direct result of photolysis by UV photons, or
indirectly the result of photolysis reactions that produce radicals
that rapidly react with other species. Cool stars emit longer
wavelength radiation, leading to a lower amount of energy in the UV
region responsible for many photolytic reactions, and correspondingly
slower photolysis rates. However, active M dwarfs such as AD Leonis
(AD Leo) emit significant energy fluxes at wavelengths shortward of
200 nm, allowing a subset of photochemical reactions to proceed at a
rate comparable to that around warmer stars. As a result, the
photolysis reactions caused by these photons may proceed at Earth-like
rates on planets around cool stars.

\begin{figure}[]
\plotone{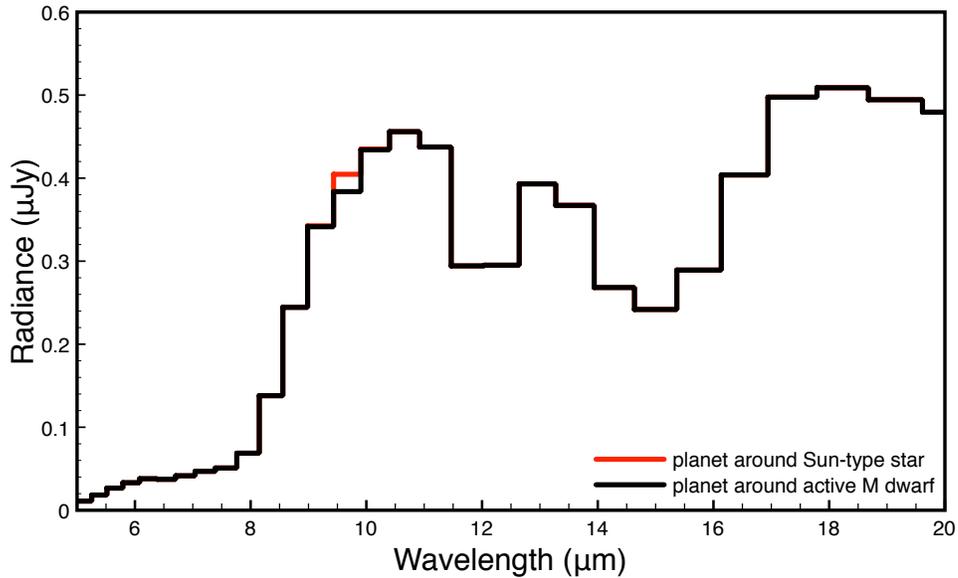}
\caption{Model spectra of planets orbiting different stars. The black 
curve is the predicted reflection spectrum from a hypothetical planet 
orbiting AD Leo with about the resolution expected for
{\it Terrestrial Planet Finder} missions. The red 
curve (see online version) shows the same, but with the planet orbiting the Sun. The difference, which is due to the abiotic build-up of O$_3$, is at the limits
of detectability. Different planetary conditions and stellar parameters may
make this discrepancy larger.}
\label{fig:spectra}
\end{figure}

As an example we consider the effects of the SEDs of the Sun and AD
Leo on a hypothetical planet orbiting in the HZ (see
\citet{Segura2005}). The abundance of O$_2$ and O$_3$ in an atmosphere
is a function of photolysis rates, and hence different SEDs could
influence the abundance of these important biomarkers in a planetary
atmosphere. On hypothetical planets around AD Leo, short-UV radiation
from flares can lead to photolysis of O$_2$, H$_2$O, and CO$_2$,
liberating O atoms. These O atoms can then react with O$_2$ to form
O$_3$. However, because O$_3$ photolysis occurs at longer wavelengths
then O$_2$ photolysis, and these longer wavelengths are relatively
scarce in all M dwarf SEDs, the destruction of the O$_3$ will be
slower. {\it This could lead to build-up of atmospheric O$_3$ from
processes that are photochemical and not biological.}

The net impact of the influence of SEDs on a planetary spectrum can be
seen in Fig.~\ref{fig:spectra}. The black line shows results from a
coupling of our photochemistry model \citep{Pavlov2001} with our
line-by-line spectral model \citep{Crisp1997}. This particular
simulation is of a spectrum from an organic-rich planet without
biological O$_2$ production in the HZ of AD Leo. The red line
represents our simulations of the same planet in the habitable zone of
the Sun. The offset between these two lines results from AD Leo's lack
of emission of photons with wavelengths in the range 200--800 nm,
which destroy O$_3$. This result demonstrates how ignoring the stellar
context of a planetary environment could lead to a false positive for
life. The SEDs of M stars likely lead to biosignatures of their
inhabited planets that are qualitatively different than those expected
from inhabited planets orbiting F, G, and K stars.

\section{Tidal Constraints on Habitability}

Terrestrial planets orbiting close to their host stars may be deformed
by the gradient of the gravitational force across their diameters. The
tidal bulge raised on the planet will generally not be aligned with
the line between the gravitational centers of the two bodies as long
as 1) the orbit is eccentric ($e~\neq~0$), or 2) the rotational period
is different from its orbital period
($P_\mathrm{p}^\mathrm{rot.}~\neq~P_\mathrm{p}^\mathrm{orb.}$), or 3)
the spin has an obliquity with respect to the orbital plane
($\psi_\mathrm{p}~\neq~0$).  Although gravity tries to align the
bulge, friction within he body resists, resulting in ``tidal
heating.''  Conservation of energy and angular momentum forces the
planet's semi-major axis $a$, $e$, $P_\mathrm{p}^\mathrm{rot.}$, and
$\psi_\mathrm{p}$ to steadily change.

Initially tides drive $\psi_\mathrm{p}~\rightarrow~0$ (``tilt
erosion''), and $P_\mathrm{p}^\mathrm{rot.}$ evolves towards the
``equilibrium rotation'' $P_\mathrm{p}^\mathrm{equ.}$. When
$\psi_\mathrm{0}~\approx~0$, the planet will not experience seasons,
i.e. over the course of an orbit, the insolation distribution on the
planet will not vary. If
$P_\mathrm{p}^\mathrm{rot.}~=~P_\mathrm{p}^\mathrm{equ.}$, one
hemisphere of the planet will permanently be irradiated by the star,
while the night side will freeze, which may prevent global
habitability \citep{Joshi2003}.  Moreover, ``tidal heating'' in the
planet may cause global volcanism or rapid resurfacing as
observed in the Solar System on Io.

\begin{figure}[!ht]
\plotone{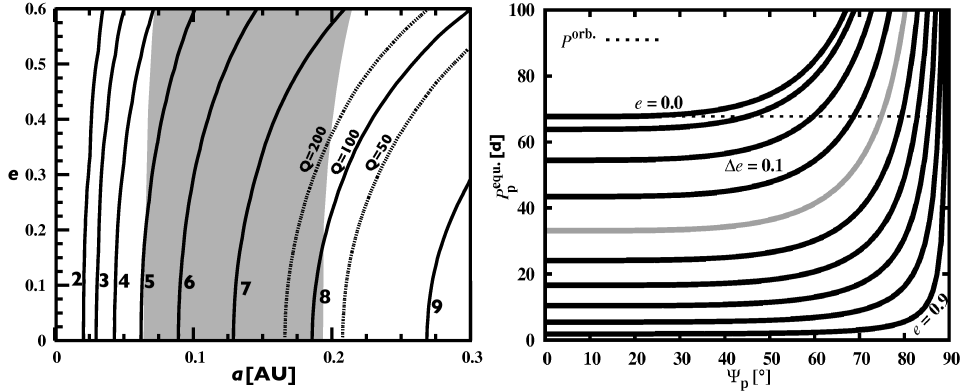}
\caption{\textit{Left:} Tilt erosion times in units of
$\log(t_\mathrm{ero}/\mathrm{yr})$ for an Earth-mass planet orbiting
a $0.25\,M_\odot$ star. The HZ is shaded in grey. \textit{Right:}
Equilibrium rotation period of Gl~581\,d as a function of obliquity
$\psi_\mathrm{p}$ for different values of $e$. The observed
$e~=~0.38~\pm~0.09$ is close to the grey line for $e~=~0.4$. The orbital
period, 67 days, is marked with a dashed line.}
\label{fig:erosion_Gl581d}
\end{figure}

We define the ``tilt erosion time'' $t_\mathrm{ero}$ to be the time
tides require to decrease an initial Earth-like obliquity of
$\psi_\mathrm{p}~=~23.5^\circ$ to $5^\circ$, which depends on the
initial $a$, $e$, and $P_\mathrm{p}^\mathrm{rot.}$. In the left panel
of Fig.~\ref{fig:erosion_Gl581d}, $t_\mathrm{ero}$ is projected on to
the $a$-$e$ plane, as calculated with the tidal model of
\citet{Leconte2010}. The tidal time lag of the planet
$\tau_\mathrm{p}$, the interval between the passage of the perturber
and the tidal bulge, is scaled by
$\tau_\mathrm{p}~=~638\,\mathrm{s}~\times~Q_\oplus/Q_\mathrm{p}$ to
fit the Earth's time lag $\tau_\oplus$ and dissipation value
$Q_\oplus$ \citep{NeronDeSurgyLaskar1997}. $Q_\mathrm{p}~=~100$ and an
initial rotation period $P_\mathrm{p}^\mathrm{rot.}~=~1$\,d are
assumed. An error estimate for $Q_\mathrm{p}$ of a factor 2 is indicated
with dashed lines. The test planet has one Earth-mass and orbits a
$0.25\,M_\odot$ star. The HZ of \citet{Barnes2008} is shaded in
grey. Obviously, planets in the HZ experience
$t_\mathrm{ero}~<~0.1$\,Gyr. For lower stellar masses,
$t_\mathrm{ero}~\ll~0.1$\,Gyr for terrestrial planets in the HZ
\citep{Heller_sub}. For terrestrial planets in highly eccentric orbits
in the HZ, tilt erosion can occur within 10\,Gyr for stellar masses as
large as $1\,M_\odot$.

The tidal equilibrium rotation period $P_\mathrm{p}^\mathrm{equ.}$ is
a function of both $e$ and $\psi_\mathrm{p}$ \citep{Hut1981}. As an
example, the right panel of Fig.~\ref{fig:erosion_Gl581d} shows
$P_\mathrm{p}^\mathrm{equ.}$ for the Super-Earth Gl 581\,d projected
onto the $e$-$\psi_\mathrm{p}$ plane. Observations \citep{Mayor2009}
provide $e~=~0.38~\pm~0.09$ (grey line), while $\psi_\mathrm{p}$ is
not known. At an age of $\gtrsim~2$\,Gyr
\citep{Bonfils2005}, an initial Earth-like obliquity of the
planet is already eroded \citep{Heller_sub}. For
$\psi_\mathrm{p}~\lesssim~40^\circ$, then
$P_\mathrm{p}^\mathrm{rot.}~\approx~P_\mathrm{p}^\mathrm{orb.}/2$.

``Orbital shrinking'' ($a~\rightarrow~0$) may pull an adequately
irradiated planet out of the HZ
\citep{Barnes2008}. Thus, planets observed outside the HZ
might have been habitable once in the past or become habitable in the
future. Similarly planets currently in the HZ may have been
inhospitable earlier. Terrestrial planets in the HZ of stars with
masses $\leq~0.25\,M_\odot$ undergo significant tidal heating,
potentially causing global volcanism
\citep{Jackson2008c,Barnes2009,Heller_sub}, possibly rendering such
planets uninhabitable. The consideration of tidal processes affects
the concept of the habitable zone. Tilt erosion and equilibrium
rotation need to be considered by atmospheric scientists, while
orbital shrinking and tidal heating picture scenarios for geologists.

\section{Evaporated/ing Cores of Gas Giants}

Several hot Jupiters, e.g. HD 209458 b,
show evidence of mass loss from their atmospheres
\citep{VidalMadjar2004}. Many M dwarfs are far more active than the
typical known planet-hosting stars. Hence, gas giants around these stars
may also be losing significant mass. If these planets could be
stripped of all their gas, a rocky/icy core could be left behind
\citep{Raymond2008}. Furthermore, since {\it in situ} formation of
large terrestrial planets (larger than Mars) appears challenging
\citep{Raymond2007}, {\it detectable} rocky planets in the HZ of M
stars may have followed just such an evolution. The core
accretion model of planet formation \citep{Pollack1996,Lissauer2009}
posits that the cores of giant planets formed beyond the ``snow line''
(the region of a protoplanetary disk which is cold enough to permit
the formation of water ice), and hence we may expect such cores to be
volatile-rich. This section explores the possibility that
terrestrial planets in the HZ of M dwarfs could be the
remnant cores of ice or gas giants.

Atmospheric mass loss is most tightly coupled to the extreme
ultraviolet flux, $F_{XUV}$, incident on a planet
\citep{Baraffe2004}. On FGK stars, $F_{XUV}$ drops over a timescale of
Gyr \citep{Ribas2005}. For M dwarfs, stellar activity often produces
XUV photons, but also decreases with time \citep{West2008}. Therefore,
for a constant orbit, we expect the mass loss rate to decrease with
time.

The complete removal of a gas giant's atmosphere would likely leave
behind a core with a mass of perhaps several $\textrm{M}_\oplus$
\citep{Baraffe2004,Raymond2008}. On the other hand, some studies argue
that the observations of HD 209458 b do not imply significant loss of
mass \citep{BenJaffel2007}, and a theoretical
study by \citet{MurrayClay2009} suggested that complete
evaporation of a gas giant's atmosphere is unlikely. These competing
hypotheses may now be testable. With the detection capabilities of the
\emph{Kepler} and \emph{CoRoT} missions, rocky planets arising from a
variety of histories may be detected. 

In addition to mass loss, tides will play an important and
interrelated role in the evolution of an evaporating gas or ice
giant. \citet{Jackson2010} showed that the coupling of mass loss and
orbital evolution may have played a significant role in CoRoT-7 b's
history. Several important feedbacks between mass loss and tidal
evolution are possible. As $a$ decreases, tides will accelerate
orbital decay and mass loss. However, as mass decreases, orbits can
decay more slowly. Orbital decay can occur on a timescale of Gyr,
similar to the timescale for $F_{XUV}$ for M dwarfs to
diminish. Therefore, as tides pull a planet in, the mass loss
increases due to proximity, but decreases due to less flux, but as
mass is lost, tidal evolution slows.

Although \citet{Jackson2010} explored mass loss for CoRoT-7 b, the
analogous problem for M dwarfs has yet to be tackled, in part because
$F_{XUV}$ is poorly constrained. In particular for M dwarfs, stellar
activity does not evolve smoothly, as it is often punctuated by strong
outbursts that may drive fast mass loss. Nonetheless, the example of
CoRoT-7 b suggests that similar processes may be important for planets
orbiting M dwarfs. As such planets are discovered, determination of
their histories will be critical, as evaporated cores may not even be
habitable.

\section{Magnetic Shielding of Exo-Earths in the Habitable Zones of M Dwarfs}

Many planets in the HZs of M dwarfs will be exposed to denser stellar
winds and be tidally locked. As a consequence, the exoplanet community
initially reached the consensus that the slow rotation of the planets
will prevent the development of magnetic fields strong enough to
shield surface life. In this section, we show that a planet in the HZ
of an M dwarf, even if rotating very slowly, can have stronger
magnetic shielding than previously thought.

Although planets in the HZs of M dwarfs do not necessarily rotate
synchronously due to tides (see $\S$~4, \citet{Correia2008} or
\citet{Barnes2010b}), tidal locking still leads to slow rotation,
except for the very latest M dwarfs and/or large
eccentricities. Planetary scientists have been working for decades on
models to reproduce the magnetic moments $\mathcal{M}$ of planets and
satellites in our Solar System. The best model generated to date, in
the sense of reproducing the measured magnetic moments of most objects
in the Solar System, is by \citet{OlsonChristensen2006}.

We applied that model to the case of hypothetical exo-Earths (with masses up
to 12 $\textrm{M}_\oplus$), to determine the strengths of their magnetic
fields. We assume their interiors are stratified in two separate
layers, a mantle and a core. We also assume that the planets have thin
atmosphere and ocean/crust layers which account only for 1\% of the
planetary radius.  We use two different chemical compositions for each
layer: a pure iron core and an iron alloy core, containing 10 mole \% S, and for the mantle we use pure olivine and
perovskite+ferropericlase compositions, which resemble, respectively,
the upper and lower mantle compositions of the Earth.

We implemented all possible combinations of those model layers into
the standard equations of planetary structure to obtain the density
profile of the planets and from those density profiles computed the
parameters needed to determine their magnetic moment, i.e. the radius
of the core, the average bulk density of the convective zone, and its
thickness.

\begin{figure}[!ht]
\plotone{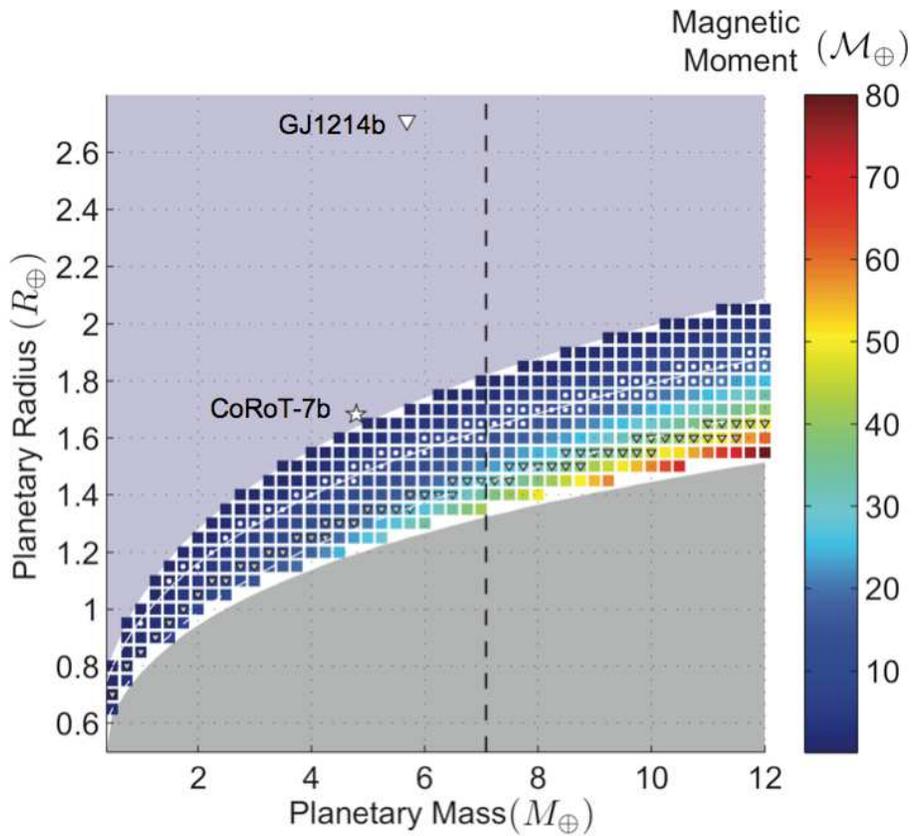}
\caption{\label{fig:magnetic}Magnetic moment model estimates for planets 
with masses up to 12 $\textrm{M}_\oplus$, a pure iron core, and 
perovskite+ferropericlase mantle compositions. The color scale (see
online version) on the right 
corresponds to magnetic moment values between 0 and 80 times that of
the Earth. The region below the colored points corresponds to 
planets made out of core materials denser than iron, while the region 
above corresponds to planets with radii too large, and therefore too low 
density, to have a core capable of generating a magnetic field. The 
triangle in the upper edge of the plot corresponds to GJ 1214 b, and the 
star symbol corresponds to CoRoT-7 b \citep{Leger2009}. For the core 
and mantle compositions in these models, neither of those two planets will 
have a magnetic field, but this result can change in the case of
CoRoT-7 b 
for slightly different interior chemical compositions.}
\end{figure}

The magnitude of the magnetic moment of a planet varies with density
and core-mantle composition. Pure iron core and
perovskite+ferropericlase mantle models result in the smallest
planetary core radius and therefore the smallest moment, but even in
this case the value for Earth-sized planets like those in our
simulations will still be at least $0.4\mathcal{M}_\oplus$,
independent of the planets' rotation rates.  All other models produce
stronger dipoles.  Fig.~\ref{fig:magnetic} summarizes our findings for
a grid of simulated planetary mass and radius values of exo-Earths
with masses up to 12 $\textrm{M}_\oplus$, assuming the mantle and core
compositions that gave the weakest magnetic moment, as described
above. Therefore, the values in the figure are lower limits to the
expected magnetic moment strengths.

Planets with larger masses and smaller radii are more likely to have
larger dipole moments and stronger magnetic fields at the surface,
because denser planets tend to have larger cores. The main conclusions
are: 1) \emph{the magnetic moment of a planet does not depend on its
rotation rate}; 2) the magnetic moment depends instead on its mass and
size, its chemical composition, and the efficiency of convection in
its interior; and 3) any terrestrial planet up to a few Earth masses
in the HZ of an M-dwarf might have a strong enough magnetic field to
shield its atmosphere and surface.

Notice, however, that these models do not account for changes in the
thickness of the convective zone or the convective flux. Also, planets
under extreme conditions, i.e. highly inhomogeneous heating or very
strong stellar winds, will undoubtedly have their magnetic fields
affected.

\section{Revisiting the Habitable Zone:  Summary of the NAI Workshop Discussions}

To take stock of the current state of the field of planetary
habitability, the NASA Astrobiology Institute's Habitability and
Biosignatures Focus group organized a workshop in Seattle, WA on Aug
3--5, 2010.  This workshop gathered together 38 scientists from the
fields of biology, geology, atmospheric science, ecology and
astronomy.  The primary goal of the workshop was to identify, review
and prioritize planetary and stellar characteristics that affect
habitability, and to provide an interdisciplinary synthesis of these
developments in our understanding of planetary habitability.  As a
secondary goal, the workshop initiated a discussion on the development
of a multi-parameter means of assessing the likelihood of extrasolar
planet habitability.  In this section we will briefly describe
workshop highlights in the major topic areas covered. The full report
will be available in 2011 (Meadows et al., in prep.).

{\bf Life's Requirements:} The recently-published NRC report on
``The Limits of Organic Life in Planetary
Systems''\footnote{http://www.nap.edu/catalog.php?record\_id=11919}
finds that life requires a scaffolding element which can form covalent
bonds with other elements that are relatively easily broken.  Weaker,
non-covalent electrostatic bonds are also crucial, as they are
required to maintain the 3-D structure of proteins.  As life's
molecules require both covalent and non-covalent bonds to function,
this strongly favors a polar solvent like water, and much of our
discussion in this session was on scientific arguments for water as
the most likely solvent for life. Energy is also required, but in the
HZ of \citet{Kasting1993}, stellar photons provide an essentially
limitless supply. More speculatively, the boundaries of an HZ may be
a function of the metabolic pathways utilized on a particular planet.

{\bf Stellar Radiative Effects:} The star drives a planet's
climate, but can also negatively impact habitability by subjecting the
planet to high-energy radiation and particles. Modeling work on the
effect of flares on Earth-like planets orbiting in the HZ
of M dwarfs find that this latter effect may be mitigated if the planet
has an existing ozone layer, and if flares are infrequent
\citep{Segura2010}. UV radiation from the flare has little effect on
O$_3$, and it is the chemistry associated with the proton flux from
the flare that produces the most damage.  However, DNA-damaging UVB
flux reaching the surface at the peak of the flare was only 1.2 times
Earth's level.

{\bf Planetary System Architecture:} Discussion in this session
emphasized the importance of planetary rotation rate, obliquity and
eccentricity, characteristics that are often extremely difficult to
constrain observationally.  Tidal locking with synchronous rotation
can also lead to a 0 or 180 degree obliquity (planetary pole
perpendicular to the orbital plane) such that the planet's poles do
not experience seasonal melting, see $\S$~3.  This state can lead to
runaway glaciation as the poles freeze and planetary
albedo steadily increases. High obliquity and moderate eccentricity can push out the
outer edge of the HZ \citep{Spiegel2010}. Tidal heating can also be so
strong as to render a planet in the HZ uninhabitable
\citep{Jackson2008c,Barnes2009}.

{\bf Forming Habitable Planets:} Highlights of this session
included a discussion of water and carbon worlds.  Radioactive heating
from supernova-sourced $^{26}$Al can deplete water in planetesimals,
leading to drier planets. Without $^{26}$Al, Earth might have formed
with 30--50 times as much water resulting in oceans 400 km deep.
This ``waterworld'' might be uninhabitable however, as the overlying
pressure could form an ice layer on the ocean floor, cutting off
nutrient communication with the rocky interior. Planet formation and
disk-chemistry modeling suggests that 1 in 3 planetary systems has a
C/O ratio higher than about 0.8 and may form SiC as the principle
planetary constituent \citet{Bond2010}.

{\bf Planetary Characteristics:} This session 
included an in depth discussion of planetary tectonic regimes.
Internal energy is needed to drive convection for plates of a
particular strength and planets may go through different tectonic
regimes, including plate tectonics as we know it, but also episodic
overturns and stagnant lid regimes.  Even the Earth is likely to only
have had smooth plate tectonics for the last billion years
\citep{Condie1998}.  This could have ramifications for the
carbonate-silicate cycle which may have buffered the Earth's climate
for the past 4 Gyr \citep{Walker1981}. A Dune-like world, with only
1--10\% of the Earth's water abundance, could both cool more
efficiently at the inner HZ edge, and avoid an ice-albedo feedback at
the outer edge, and hence may have a much broader HZ than predicted by
\citet{Kasting1993}.

{\bf Detecting Habitability:} This session included discussions
of detecting distant oceans.  \citet{Robinson2010} have used a
realistic 3-D spectral model of the Earth to show that photometric
observations at specific extrasolar planetary phases could be used to
discriminate between planets with and without significant bodies of
liquid on their surfaces.  Polarization signals also permit the
detection of surface liquids, and the peak in polarization percentage
may give clues to atmospheric thickness.

\section{Conclusions}

Less than one month after the Cool Stars XVI meeting, \citet{Vogt2010}
reported the RV detection of a potentially rocky planet ($m_p \ge 3
\textrm{M}_\oplus$) orbiting in the HZ of the M3 star Gl 581 in a nearly
circular orbit. If confirmed, this planet is the first discovered near
the middle of the HZ of a main sequence star, and, as expected, that
star is an M dwarf. So how does this planet measure up in terms of
potential habitability? Vogt et al. note that the host star is
extremely quiescent, to the point that they cannot really detect any
jitter, and hence stellar activity is not currently an issue for the
planet. However, the star is at least a few Gyr old
\citep{Bonfils2005}, hence we cannot exclude the possibility that fatal
harm was done to a potential biosphere in the past. The planet is
tidally locked, or in a spin-orbit resonance \citep{Heller_sub}, but
we do not know its rotation rate. The planet's mass is large enough
that it can sustain tectonic activity for 10 Gyr, assuming it formed
with a similar ratio of radiogenic isotopes as the Earth (tidal
heating is minimal, even for $e=0.2$). Therefore, from the RV data, we
can not discern any major issue impeding habitability. Unfortunately,
though, remote detection of its biosphere is not possible for the
foreseeable future, as at 0.15 AU from its host star, reflection
spectra will be unavailable from any currently planned space mission.

In spite of early skepticism, planets orbiting M dwarfs can be
inhabited. Although many issues have been identified, such as tidal
locking and atmospheric removal, more careful modeling has shown that
these phenomena are not so deadly. While improvements
in our knowledge via modeling will continue, transmission spectra of planets
transiting M dwarfs will provide, at least for the next decade, our only
observational means to directly assess the habitability of planets orbiting
cool stars.


\bibliography{Barnes_R}
\end{document}